# A SLOW CONTROL SYSTEM FOR THE GARFIELD APPARATUS

M. Giacchini, F. Gramegna, INFN, Legnaro (PD), ITALY 35020, EUROPE
S. Bertocco, SISSA, Triste, ITALY 34014, EUROPE

## Abstract

The major part of the GARFIELD apparatus electronics are monitored and set up through a slow control system, which has been developed at LNL. A software package based on Lab View has been dedicated to the setting and control of 16 channels integrated Amplifiers and Constant Fraction Discriminators. GPIB controllers and GPIB-ENET interfaces have been used for the communication between the Personal Computer and the front-end of the electronics.

## 1 INTRODUCTION

In the last decade more and more sophisticated and complex apparatuses have been built [1] to overcome modern nuclear physics requirements. These detectors (see Fig. 1).have been designed in order to both cover a large fraction of the solid angle, with a high granularity in order to study precise angular correlations between the reaction partners, and to be capable of providing good energy and charge (and/or mass) resolution for a complete identification in a wide range

More compact electronic modules have been therefore developed following the growing complexity of the apparatuses. A simplification of the hardware (no elipot, no manual control, few trimmers etc.) permits the handling of a larger number of channels per module (16-32). But the integration of modules must not bring to the lack of all those characteristics necessary for nuclear physics studies, as far as linearity, resolution, possibility of setting different values of parameters like gain, shaping time, polarity etc. The read-out and process of a large (500-1000) number of channels, with almost the same performances of the standard modules, has become a requirement.

For this reason several multi-channel analogical and digitalizing modules has been newly developed [2]. A remote control system becomes therefore indispensable to set- up the modules parameters and to keep them under control during the measurements.

We will describe in this paper the slow control system of the GARFIELD apparatus, based on a LabView platform, which has been developed for the CAEN C208 Constant Fraction Discrimination and the CAEN N568 Amplifiers [3] using the KS 3988 GPIB Crate Camac Controller [4] and C117B High Speed Caenet Camac Controller.

## 2 THE CONTROL SYSTEM

### 2.1 The design

GARFIELD [5] is a composite and multipurpose apparatus made by different detectors:
- an annular three stage telescope, divided in eight sectors along the azimuthal direction, made by Ionisation Chambers, Silicon detectors (300 μm thickness), CsI(Tl) crystals.
- a TOF system based on three Position Sensitive Parallel Plate Avalanche Counters (20x20 cm$^2$), for fragment mass measurements in dissipative collisions.
- two drift chambers, placed back-to-back with respect to the target, each of which containing about 90 micro-strip on glass gas detectors and about 90 CsI(Tl) crystals.

Close to 500 preamplifiers are used for the read out of the detectors and their signals are fed into the 16 channel shaping amplifiers CAEN mod. N568B. For each input signals three outputs are provided: two correlated linear outputs, for which shaping time, gains output polarities, pole zeroes can be set by an external control, one fast output at fixed gain and polarity. The linear signals are then fed into Analog-to-Digital Converters, based on VME/FAIR bus [2], while the fast output has to be sent to the 16 channel CAEN Constant Fraction Discrimination Mod. C208.

A good and efficient control system is therefore necessary to handle the high number of channels: a LabView application has been therefore developed for this purpose.

The experimental environment is distributed as follows: the apparatus is located in the experimental area, in a scattering chamber under vacuum. All the electronics (up to the acquisition front-end) is mounted in the experimental area , close to the scattering chamber; the "acquisition data" room is connected to the experimental area via Ethernet, both from the acquisition system point of view (Optical fibers), and from the slow control system (coaxial cables).

## 2.2 Network Idea

The control system has been developed taking also into account the possibilities offered by a network environment, like for example the utilization of Internet Tools to display the status of the system and to notify possible problems and faults.

The system controls about 500 channels for both the Constant Fraction Discrimination channels and the Amplifiers and it runs on a Window 2000 PC, connected to the hardware in the experimental area through a GPIB-ENET interface. This interface transparently handles the data transfer between an Ethernet-based TCP/IP host and the GPIB. Through GPIB-ENET multiple hosts can share a set of GPIB instruments or a single host can control several GPIB systems. The GPIB-ENET converts a computer equipped with an NI-488.2 driver and an Ethernet port into a GPIB Talker/Listener/controller. The Maximum GPIB transfer rate is 100 kbytes/s. The limited rate is due to the fact that we are using a general LAN of the institute, and not a reserved one. In this case GPIB-ENET is used to control the KS 3988 Crate Camac Controller from a PC.

The control software has been developed using LabView, by National Instruments [6]. LabView is a graphical programming environment developed for data acquisition, control data analysis, data presentation. The graphical programming method used is completely flexible, becoming very close to a powerful programming language, without reaching the same difficulty and complexity. A basic library of graphical instruments, which permits to manage GPIB communication quite easily, is also provided by LabView

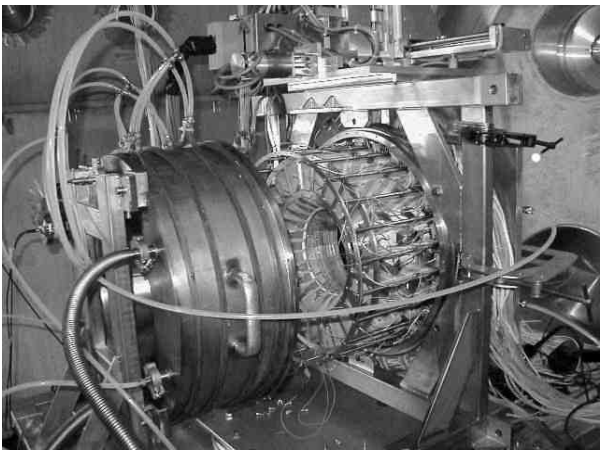

Figure 1: Detector

## 2.3 Start Up

A general set-up of all the parameters is required at the beginning of each experiment: this is a quite slow action because it concerns all the channels and almost all the settable parameters. This action is an initialization necessary to be sure some standard parameters will be automatically set (for example the very last parameters used at the end of a previous run).

## 2.4 The data

The data stored in a data base file are sent to the module front-end and a comparison checking the conformity between the parameter values in the data base and the setting in the module is performed: the status column gives out an error if differences are found. Problems can derive due to network bad transmission or malfunctioning of individual channels.

## 2.5 The GUI

The control system developed on these bases allows the handling and check of the systems parameters through a Graphical User Interface (GUI) (see Fig. 2).

The GUI can then provide these control functions:

- *Set-up of parameters*, chosen by the user, for individual channels or for groups of channels. Different channels can be grouped from the users, depending on the experimental set-up (grouping by kind of detectors, sectors or rings of the apparatus etc.);
- *storage* of the set parameters in an Excel spreadsheet used like a data-base;
- *check of the current status* of the apparatus: reading of the parameters values in the modules via TCP/IP and comparison between read data and data stored in the database.

The control system provides moreover some Internet based tools to monitor the module status and to perform the alarm notification.

The status monitor service allows the display of the updated database spreadsheet in a Web Browser, using an HTTP server. It converts the database file into a HTML file, exploiting the LabView Internet toolkit and it upgrades the file at each significant event. This file can be watched by means of a Web Browser, which making use of a java applet, reloads the file periodically. If the monitoring is performed using Internet Explorer, loading the file an Informative Microsoft Agent appears.

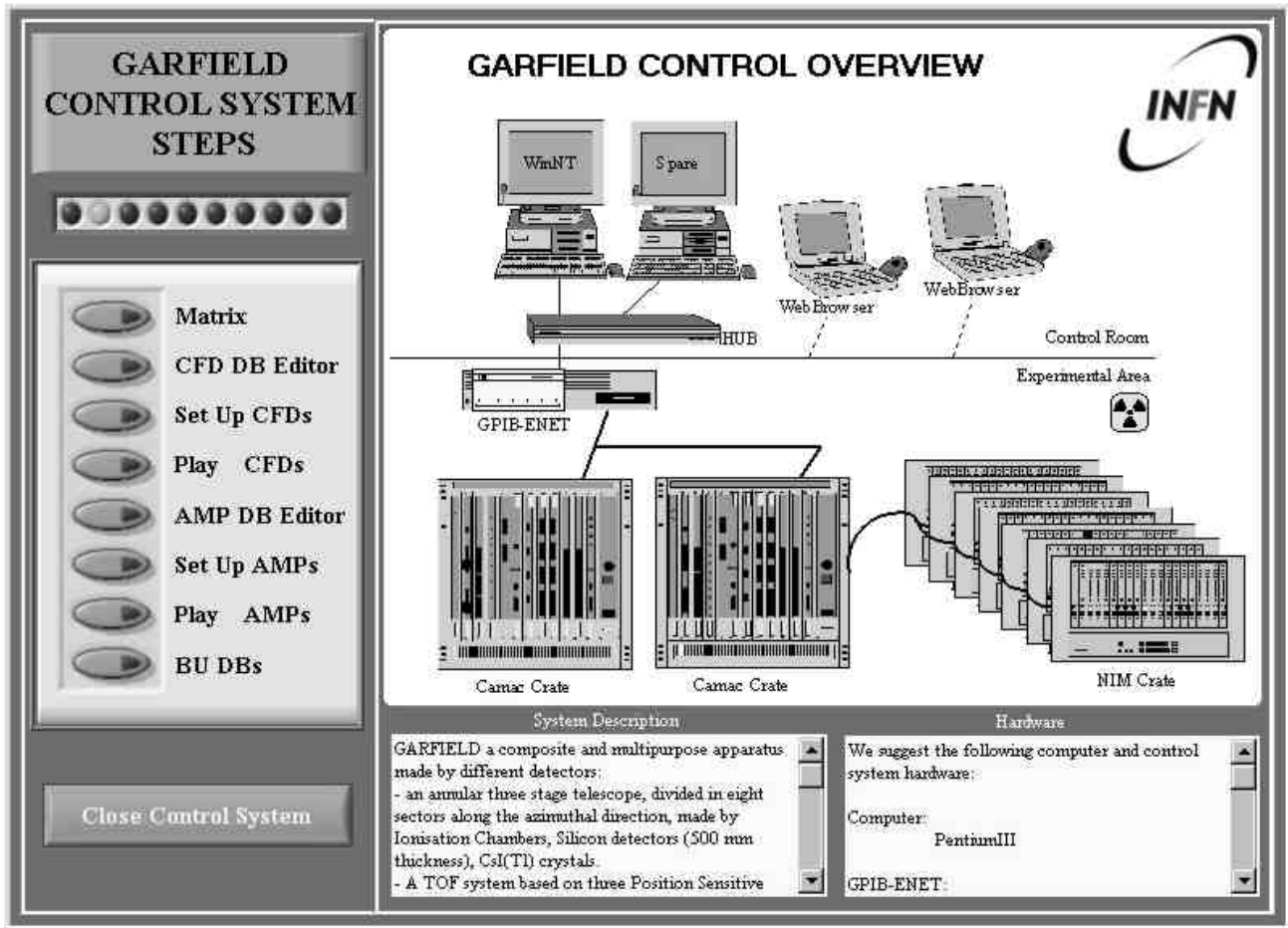

Figure 1: The GUI

This Agent can notify alarms by mean of a speech output. Alarms can also be notified by an e-mail notification of failure performed by using the LabView Internet toolkit.

## 2.6 Conclusions

A slow control system for the Garfield electronics has been developed, to set up and monitor Amplifiers and Constant Fraction Discriminators parameters. The architecture of the control was designed to be easily applied to different experiments. Some future developments are in progress with regards to High Voltages systems. Lab View platform demonstrates to be a powerful graphical programming environment, which permits a wide field of applications.

## REFERENCES

[1] F. Gramegna et al. A389(1997)474, R.T. De Souza et al. NIM A295(1990)109,

I. Iori et al. NIM A325(1993)458, J. Pouthas et al. NIM A357(1995)418

[2] R.Bassini et al NIM A305 (1991) 449, R. Bassini et al IEEE Symposium S. Fe (1991) , A. Ordine IEEE Trans. On Nucl Scien. 45(3)1998

[3] CAEN Catalogue

[4] KinetikSystemCorporation Catalogue

[5] GARFIELD:a General ARray for Fragment Identification and for Emitted Light particles in Dissipative collisions – to be published

[6] http:/www.ni.com

[7] 'Costant fraction discriminators control for Garfield' - LNL-INFN(REP)- 160/00